# HEAPABLE SEQUENCES AND SUBSEQUENCES

JOHN BYERS[1], BRENT HEERINGA[2], MICHAEL MITZENMACHER[3], AND GEORGIOS ZERVAS[1]

ABSTRACT. Let us call a sequence of numbers heapable if they can be sequentially inserted to form a binary tree with the heap property, where each insertion subsequent to the first occurs at a leaf of the tree, i.e. below a previously placed number. In this paper we consider a variety of problems related to heapable sequences and subsequences that do not appear to have been studied previously. Our motivation for introducing these concepts is two-fold. First, such problems correspond to natural extensions of the well-known secretary problem for hiring an organization with a hierarchical structure. Second, from a purely combinatorial perspective, our problems are interesting variations on similar longest increasing subsequence problems, a problem paradigm that has led to many deep mathematical connections.

We provide several basic results. We obtain an efficient algorithm for determining the heapability of a sequence, and also prove that the question of whether a sequence can be arranged in a *complete* binary heap is NP-hard. Regarding subsequences we show that, with high probability, the longest heapable subsequence of a random permutation of $n$ numbers has length $(1 - o(1))n$, and a subsequence of length $(1 - o(1))n$ can in fact be found online with high probability. We similarly show that for a random permutation a subsequence that yields a complete heap of size $\alpha n$ for a constant $\alpha$ can be found with high probability. Our work highlights the interesting structure underlying this class of subsequence problems, and we leave many further interesting variations open for future work.

[1]DEPARTMENT OF COMPUTER SCIENCE, BOSTON UNIVERSITY
[2]DEPARTMENT OF COMPUTER SCIENCE, WILLIAMS COLLEGE
[3]DEPARTMENT OF COMPUTER SCIENCE, HARVARD UNIVERSITY
*E-mail addresses*: byers@cs.bu.edu, heeringa@cs.williams.edu, michaelm@eecs.harvard.edu, zg@bu.edu.
John Byers and Georgios Zervas are supported in part by Adverplex, Inc. and by NSF grant CNS-0520166.
Brent Heeringa is supported by NSF grant IIS-08125414.
Michael Mitzenmacher is supported by NSF grants CCF-0915922 and IIS-0964473, and by grants from Yahoo! Research and Google Research.

0

## 1. Introduction

The study of longest increasing subsequences is a fundamental combinatorial problem, and such sequences have been the focus of hundreds of papers spanning decades. In this paper, we consider a natural, new variation on the theme. Our main question revolves around the problem of finding the *longest heapable subsequence*. Formal definitions are given in Section 2, but intuitively: a sequence is heapable if the elements can be sequentially placed one at a time to form a binary tree with the heap property, with the first element being placed at the root and every subsequent element being placed as the child of some previously placed element. For example, the sequence $1, 3, 5, 2, 4$ is heaple, but $1, 5, 3, 2, 4$ is not. The longest heapable subsequence of a sequence then has the obvious meaning. (Recall that a subsequence need not be contiguous within the sequence.)

Our original motivation for examining such problems stems from considering variations on the well-known secretary problem [5, 6] where the hiring is not for a single employee but for an organization. For example, Broder et al. [3] consider an online hiring rule where a new employee can only be hired if they are better than all previous employees according to some scoring or ranking mechanism. In this scenario, with low ranks being better, employees form a decreasing subsequence that is chosen online. They also consider rules such as a new employee must be better than the median current employee, and consider the corresponding growth rate of the organization.

A setting considered in this paper corresponds to, arguably, a more realistic scenario where hiring is done in order to fill positions in a given organization chart, where we focus on the case of a complete binary tree. A node corresponds to the direct supervisor of its children, and we assume the following reasonable hiring restriction: a boss must have a higher rank than their reporting employees.[1] A natural question is how to best hire in such a setting. Note that, in this case, our subsequence of hires is not only heapable, but the heap has a specific associated shape. As another variation our organization tree may not have a fixed shape, but must simply correspond to a binary tree with the heap property—at most two direct reports per boss, with the boss having a higher rank.

We believe that even without this motivation, the combinatorial questions of heapable sequences and subsequences are compelling in their own right. Indeed, while the various hiring problems correspond to online versions of the problem, from a combinatorial standpoint, offline variations of the problem are worth studying as well. Once we open the door to this type of problem, there are many fundamental questions that can be asked, such as:

- Is there an efficient algorithm for determining if a sequence is heapable?
- Is there an efficient algorithm for finding the longest heapable subsequence?
- What is the probability that a random permutation is heapable?
- What is the expected length and size distribution of the longest heapable subsequence of a random permutation?

We have answered some, but not all, of these questions, and have considered several others that we describe here. We view our paper as a first step that naturally leads to many questions that can be considered in future work.

1.1. **Overview of Results.** We begin with heapable *sequences*, giving a natural greedy algorithm that decides whether a given sequence of length $n$ is heapable using $O(n)$ ordered dictionary operations. Unfortunately, when we place further restrictions on the shape of the heap, such as insisting on a complete binary tree, determining heapability becomes NP-hard. Our reduction involves gadgets that force subsequences to be heaped into specific shapes which we exploit in delicate ways. However when the input sequence is restricted to 0-1 the problem again becomes tractable and we give a linear-time algorithm to solve it. This case corresponds naturally to the scenario where candidates are rated as either *strong* or *weak* and strong candidates will only work for other strong candidates (weak candidates are happy to work for whomever).

---

[1]We do not claim that this *always* happens in the real world.

Turning to heapable *subsequences*, we show that with high probability, the length of the longest heapable subsequence in a random permutation is $(1 - o(1))n$. This result also holds in the online setting where elements are drawn uniformly at random from the unit interval, or even when we only know the ranking of a candidate relative to the previous candidates. In the case when we restrict the shape of the tree to complete binary trees, we show that the longest heapable subsequence has length linear in $n$ with high probability in both the offline and online settings. In all cases our results are constructive, so they provide natural hiring strategies in both the online and offline settings. Throughout the paper, we conduct Monte Carlo simulations to investigate scaling properties of heapable subsequences at a finer granularity than our current analyses enable. Finally, we discuss several attractive open problems.

1.2. **Previous Work.** The problems we consider are naturally related to the well-known longest increasing subsequence problem. As there are hundreds of papers on this topic, we refer the reader to the excellent surveys [1, 8] for background.

We briefly summarize some of the important results in this area that we make use of in this paper. In what follows, we use LIS for longest increasing subsequence and LDS for longest decreasing subsequence. Among the most basic results is that every sequence of $n^2 + 1$ distinct numbers has either an LIS or LDS of length at least $n+1$ [4, 8]. An elegant way to see this is by greedy patience sorting [1]. In greedy patience sorting, the number sequence, thought of as a sequence of cards, is sequentially placed into piles. The first card starts the leftmost pile. For each subsequent card, if it is larger than the top card on every pile, it is placed on a new pile to the right of all previous piles. Otherwise, the card is placed on the top of the leftmost pile for which the top card is larger than the current card. Each pile is a decreasing subsequence, while the number of piles is the length of the LIS – the LIS is clearly at most the number of piles, and since every card in a pile has *some* smaller card in the previous pile, the LIS is at least the number of piles as well.

In the case of the LIS for a random permutation of $n$ elements, it is known that the asymptotic expected length of the LIS grows as $2\sqrt{n}$. More detail regarding the distribution and concentration results can be found in [2]. In the online setting, where one must choose whether to add an element and the goal is to obtain the longest possible increasing subsequence, there are effective strategies that obtain an asymptotic expected length of $\sqrt{2n}$. Both results also hold in the setting where instead of a random permutation, the sequence is a collection of independent, uniform random numbers from $(0, 1)$.

## 2. Definitions

Let $x = x_1, \ldots, x_n$ be a sequence of $n$ real numbers. We say $x$ is *heapable* if there exists a binary tree $T$ with $n$ nodes such that every node is labelled with exactly one element from the sequence $x$ and for every non-root node $x_i$ and its parent $x_j$, $x_j \leq x_i$ and $j < i$. Notice that $T$ serves as a witness for the heapability of $x$. We say that $x$ is *completely heapable* if $x$ is heapable and the solution $T$ is a complete binary tree.

If $T$ is a binary tree with $k$ nodes, then there are $k+1$ free slots in which to add a new number. We say that the *value* of a free child slot is the value of its parent, as this represents the minimum value that can be placed in the slot while preserving the heap property. Let $sig(T) = \langle x_1, x_2, \ldots, x_{k+1} \rangle$ be the values of the free slots of $T$ in non-decreasing sorted order. We call $sig(T)$ the *signature* of $T$. For example, heaping the sequence 1, 4, 2, 2 yields a tree with 5 slots and signature $\langle 2, 2, 2, 4, 4 \rangle$. Given two binary trees $T_1$ and $T_2$ of the same size $k$, we say that $T_1$ *dominates* $T_2$ if and only if $sig(T_1)[i] \leq sig(T_2)[i]$ for all $1 \leq i \leq k$ where $sig(T)[i]$ is the value of slot $i$ of $T$.

Now define the *depth* of a slot $i$ in $T$ to be be the depth of the parent node associated with slot $i$ of $T$. We say that $T_1$ and $T_2$ have *equivalent frontiers* if and only there is a bijection between slots of $T_1$ and slots of $T_2$ that preserves both value and depth of slots. A sequence is *uniquely heapable* if all valid solution trees for the sequence have equivalent frontiers.

Given a sequence, we say a subsequence (which need not be contiguous) is heapable with the obvious meaning, namely that the subsequence is heapable when viewed as an ordered sequence. Hence we may talk about the *longest heapable subsequence* (LHS) of a sequence, and similarly the *longest completely-heapable subsequence* (LCHS).

We also consider heapability problems on *permutations*. In this case, the input sequence is a permutation of the integers $1, \ldots, n$. For offline heapability problems, heaping an arbitrary sequence of $n$ distinct real numbers is clearly equivalent to heaping the corresponding (i.e. rank-preserving) permutation of the first $n$ integers. Here we assume the input sequence is drawn uniformly at random from the set all of $n!$ permutations on $[1, n]$. Several of our results show that given a random permutation $x$ on $[1, n]$ that the LHS or LCHS has length $f(n)$ *with high probability*, i.e. with probability $1 - o(1)$.

## 3. Heapable Sequences

### 3.1. Heapability in polynomial time.
In this section we give a simple greedy algorithm Greedy-Sig that decides whether a given input sequence is heapable using $O(n)$ ordered associative array operations, and explicitly constructs the heap when feasible.

Greedy-Sig builds a binary heap for a sequence $x = x_1, \ldots, x_n$ by sequentially adding $x_i$ as a child to the the tree $T_{i-1}$ built in the previous iteration, if such an addition is feasible. The greedy insertion rule is to add $x_i$ into the slot with the *largest* value smaller than or equal to $x_i$. To support efficient updates, Greedy-Sig also maintains the signature of the tree, $sig(T_i)$, where each element in the signature points to its associated slot in $T_i$. Insertion of $x_i$ therefore corresponds to first identifying the predecessor, PRED($x_i$), in $sig(T_{i-1})$ (if it does not exist, the sequence is not heapable). Next, $x_i$ is inserted into the corresponding slot in $T_{i-1}$, coupled with deleting PRED($x_i$) from $sig(T_{i-1})$, and inserting two copies of $x_i$, the slots for $x_i$'s children. Greedy-Sig starts with the tree $T_1 = x_1$ and iterates until it exhausts $x$ (in which case it returns $T = T_n$) or finds that the sequence is not heapable. Standard dictionary data structures supporting PRED, INSERT and DELETE require $O(\log n)$ time per operation, but we can replace each number with its rank in the sequence, and use van Emde Boas trees [9] to index the signatures, yielding an improved bound of $O(\log \log n)$ time per operation, albeit in the WORD RAM model.

**Theorem 1.** *$x$ is heapable if and only if* Greedy-Sig *returns a solution tree $T$.*

*Proof.* Let $T_1$ and $T_2$ be binary trees, each with $k$ leaves. Let $y$ be a real number such that $y \geq sig(T_2)[1]$. It is easy to see that the following claim holds.

**Claim 1.** *If $sig(T_1)$ dominates $sig(T_2)$ then $sig(T'_1)$ dominates $sig(T'_2)$ where $T'_2$ is any valid tree created by adding $y$ to $T_2$ and $T'_1$ is the tree produced by greedily adding $y$ to $T_1$.*

If Greedy-Sig returns a solution then by construction, $x$ is heapable. For the converse, let $x = x_1, \ldots, x_n$ be a heapable sequence and let $T^*$ be a solution for $x$. Since $T^*$ is a witness for $x$, it defines a sequence of trees $T^*_1, T^*_2, \ldots, T^*_n = T^*$. It follows from Claim 1 that at each iteration, the greedy tree $T_i$ strictly dominates $T^*_i$, thus Greedy-Sig correctly returns a solution. $\square$

We used Greedy-Sig to compute the probability that a random permutation of $n$ numbers is heapable as $n$ varies. The results are displayed in Figure 3.

### 3.2. Hardness of complete heapability.
We now show that the problem of deciding whether a sequence is *completely* heapable is NP-complete. First, complete heapability is in NP since a witness for $x$ is just the final tree, $T$, if one exists. To show hardness, we reduce from the NP-hard problem Exact Cover by 3-Sets which, when given a set of $n$ elements $Y = \{1, \ldots, n\}$ and a

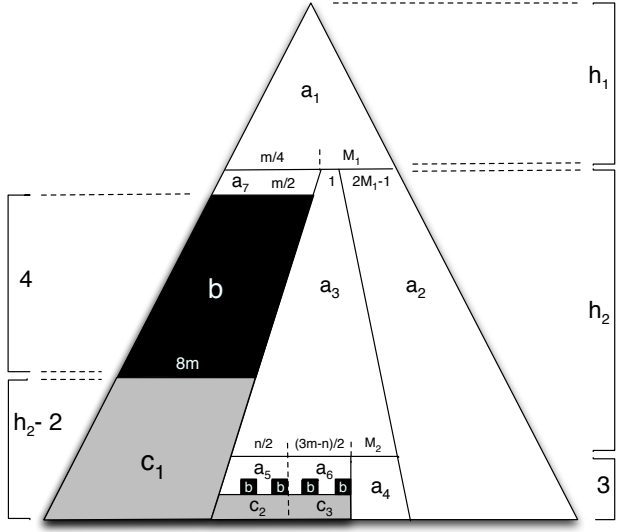

FIGURE 1. A schematic of the heap that $x$ forces. Since the prologue sequence $a_1, \ldots, a_7$ and epilogue sequence $c_1, c_2, c_3$ are uniquely heapable, the complete heapability of $x$ reduces to fitting the sequence $b$ into the black area.

FIGURE 2. An iterative definition of $\Delta(x, k, h)$.

$\Delta(x, k, h):$
1: **for** $i \leftarrow 0$ *to* $(h-1)$ **do**
2:   **for** $j \leftarrow k \cdot 2^i$ *down to* 1 **do**
3:     *print* $(x, i, j)$
4:   **end for**
5: **end for**

collection of $m$ subsets $\mathcal{C} = \{C_1, \ldots, C_m\}$ such that each $C_i \subset Y$ and $|C_i| = 3$, asks whether there exists an *exact cover* of $Y$ by $C$: a subset $C' \subset C$ such that $|C'| = n/3$ and

$$\bigcup_{C_j \in C'} C_j = Y.$$

3.2.1. *Preliminaries.* Without loss of generality, we use *triples* of real numbers in our reduction instead of a single real number and rely on lexicographic order for comparison. Our construction relies on the following set of claims that force subsequences of $x = x_1, \ldots, x_t$ to be heaped into specific shapes.

**Claim 2.** *If $x_i > x_j$ for all $j > i$ then $x$ is heapable only if $x_i$ appears as a leaf in the heap.*

*Proof.* Any child of $x_i$ must have a value $x_j \geq x_i$ with $j > i$, a contradiction. $\square$

**Claim 3.** *If $x' = x'_1, x'_2, \ldots, x'_k$ is a decreasing subsequence of $x$ then for all $x'_i$ and $x'_j$, $i \neq j$, $x'_j$ cannot appear in a subtree rooted at $x'_i$ (and vice-versa).*

*Proof.* Take such a subsequence and a pair $x'_i$ and $x'_j$. $x'_j$ succeeds $x'_i$ in the input, so $x'_i$ cannot be a descendant of $x'_j$. Also, $x'_j$ cannot be a descendant of $x'_i$ without violating the heap property. $\square$

We use claim 3 to create sequences that impose some shape on the heap. For example, consider the sequence $u = (1, 0, 2), (1, 0, 1), (1, 1, 4), \ldots, (1, 1, 1), (1, 2, 8), \ldots, (1, 2, 1)$, which, when occurring after $(1, 0, 0)$, must be heaped into two perfect binary subtrees of height 3. Since we generate sequences like $u$ often in our reduction, we use $\Delta(x, k, h)$ to denote a sequence of values of length $k(2^h - 1)$, all of the form $(x, *, *)$, that can be heaped into $k$ perfect binary trees of height $h$. Figure 2 gives an iterative definition of $\Delta$ whereby $\Delta(1, 2, 3)$ generates $u$.

**Claim 4.** *A sequence $\Delta(x, k, h)$ spans initial width at least $k$, and consumes depth at most $h$. These bounds on width and depth are also simultaneously achievable.*

*Proof.* The initial $k$ values of $\Delta(x, k, h)$ ($i = 0$ in Figure 2) are decreasing and by Claim 3, must therefore be placed at $k$ distinct leaves of the heap. The longest increasing subsequence of $\Delta(x, k, h)$ is formed by choosing one element $(x, i, *)$ for each $i$, and thus the deepest heapable subsequence of $\Delta(x, k, h)$ is $h$. To achieve these bounds tightly, simply store $\Delta(x, k, h)$ level-wise in a row of $k$ free slots. $\square$

We also define $\Gamma(x, k, h)$ to be the prefix of $\Delta(x, k, h)$ that omits the final $k$ terms, i.e. a sequence of length $k(2^h - 2)$ that can be heaped into $k$ complete binary trees with $k$ elements missing in the final level. We can now generalize Claim 3 as follows:

**Claim 5.** *If $x' = F_1(s_1, k_1, h_1), F_2(s_2, k_2, h_2), \ldots, F_t(s_t, k_t, h_t)$ is a subsequence of $x$ such that the sequence $\{s_i\}$ is decreasing and such that $F_i \in \{\Delta, \Gamma\}$ for all $i$, then for every $x_i' \in F_i(s_i, k_i, h_i), x_j' \in F_j(s_j, k_j, h_j), i \neq j$, $x_i'$ and $x_j'$ have no ancestor / descendant relationship.*

### 3.2.2. *The Reduction.*

**Theorem 2.** *Complete heapability is NP-Hard.*

*Proof.* Given an EXACT COVER BY 3-SETS instance $(Y, \mathcal{C})$ where $|Y| = n$ and $|\mathcal{C}| = m$, we construct a sequence $x = a, b, c$ of length $2^h - 1$ where $h$ is the height of the heap and $x$ is partitioned into a *prologue* sequence $a$, a *subset* sequence $b$, and an *epilogue* sequence $c$.

**Prologue sequence.** The prologue sequence $a$ consists of seven consecutive sequences $a = a_1$, $a_2$, $a_3$, $a_4$, $a_5$, $a_6$, and $a_7$:

$a_1$: $\Delta(-3, 1, h_1)$   $a_2$: $\Delta(Z, 2M_1 - 1, h_2 + 3)$         $a_3$: $\Delta(-1, 1, h_2)$         $a_4$: $\Delta(Y, M_2, 3)$
$a_5$: $\Gamma(n - \epsilon, 1, 2), \Gamma((n-1) - \epsilon, 1, 2), \ldots, \Gamma(1 - \epsilon, 1, 2)$   $a_6$: $\Gamma(0 - \epsilon, 3m - n, 2)$   $a_7$: $\Delta(-2, \frac{m}{2}, 1)$

**Epilogue sequence.** Similarly the epilogue sequence is defined to be $c = c_1, c_2, c_3$:

$c_1$: $\Delta(X, 8m, h_2 - 2)$   $c_2$: $\Delta(n, 4, 1), \Delta(n-1, 4, 1), \ldots, \Delta(1, 4, 1)$   $c_3$: $\Delta(0.1, 6m - 2n, 1)$

Taken together, the prologue and epilogue sequences enforce the following key property.

**Claim 6.** *The prologue sequence $a$ is uniquely heapable; moreover, if $x$ is completely heapable, then the epilogue sequence $c$ is uniquely heapable with respect to $a$ and $b$.*

*Proof.* By Claim 4, the sequence $a_1$ forces a complete binary tree with $N_1$ leaves. Call this tree $T_{a_1}$. Now consider the subsequence $a_2, a_3, a_7$. Since the sequence $Z, -1, -2$ is decreasing, by Claim 5, these blocks have no ancestor/descendant relationships. Moreover, since values of $a_3$ are strictly smaller than those of $a_2$ and values of $a_7$ are strictly smaller than those of $a_2 \ldots a_6$, these three blocks must all be rooted at $a_1$. Since $a_2, a_3$ and $a_7$ begin with decreasing subsequences of length $2M_1 - 1, 1$, and $m/2$ respectively, these values fill the $2 \cdot (M_1 + m/4)$ children of $a_1$, and thus the remaining levels of $a_2$ and $a_3$ are forced, also by Claim 4 (see Figure 1).

Next consider the subsequence $a_4, a_5, a_6$. At the time these values are inserted, attachment points are only available beneath $a_3$, as $a_2$ reached the bottom of the heap and remaining slots below $a_1$ are reserved for $a_7$. Since the sequence $Y, n, n-1, \ldots, 1, 0$ is decreasing, Claim 5 ensures that the components of $a_4$ through $a_6$ lie side-by-side beneath $a_3$. The construction of $a_5$ forces $n$ free slots at level $h_1 + h_2 + 2$ beneath parents of respective values $n - \epsilon, n - 1 - \epsilon, \ldots, 1 - \epsilon$. The construction of $a_6$ forces $3m - n$ free slots at that same level beneath parents of values $0 - \epsilon$. The white area of Figure 1 depicts the final shape of $a$.

As for the epilogue sequence, by Claim 2, the sequence $c_2, c_3$, as well as the final subsequence in $c_1$ must all be on the bottom row of the heap. This completely fills the bottom row of the heap (after $a$). Then by Claim 5, $c_1, c_2$ and $c_3$ have no ancestor-descendant relationship, so the rest of $c_1$ forms a contiguous trapezoid of height $h_2 - 2$ with the top row having length $8m$. The grey area of Figure 1 depicts the final shape of $c$. $\square$

This property ensures that after uniquely heaping $a$ we produce the specific shape depicted by the white area in Figure 1. Then, given that sequence $c$ is uniquely heapable with respect to $a$ and $b$, $c$ also produces a specific shape depicted by the shaded area in Figure 1. Taken together, the prologue and epilogue force sequence $b$ to be heaped into the black area of Figure 1.

The height of the heap, $h$, is defined below. Without any loss of generality, we assume $m$ is a multiple of 4 and, for convenience, define the following values

$$h_1 = \lceil \log_2(m/4 + 1) \rceil \quad N_1 = 2^{h_1} \quad M_1 = N_1 - m/4$$
$$h_2 = \lceil \log_2 3m/2 \rceil \qquad N_2 = 2^{h_2} \quad M_2 = N_2 - 3m/2.$$

Finally, let $h = h_1 + h_2 + 3$, $K = 2^h$, $L = K + 1$, $X = K + 2$, $Y = K + 2$, $Z = K + 3$ and $\epsilon$ be a small constant such that $0 < \epsilon < 1$. $Z$, $Y$, $X$, $L$ and $K$ are the 5 largest values appearing in the first position of any tuple in our sequence $x$.

Consider Figure 1 again. Sandwiched between $a_7$ and the trapezoid formed by $c_1$ is room for $m$ complete binary trees of depth 4. We call these the *tree slots*. A similar sandwich of $3m$ singleton slots is formed between $a_5, a_6$ on the top and $c_2, c_3$ on the bottom. More precisely, from the specific construction of $a$ and $c$, there are $3m - n$ *slack slots* sandwiched between $a_6$ and $c_3$ and there are $n$ *set cover slots* sandwiched between $a_5$ and $c_2$

**Claim 7.** *Each slack slot can only accept some value in the range $(0 - \epsilon, 0.1)$, and each set cover slot with parent value $i - \epsilon$ can only accept some value in the range $(i - \epsilon, i.0)$.*

*Proof.* The values in $c_3$ are strictly smaller than those in $a_5$, so they must be placed below $a_6$. Each resulting slack slot therefore has a parent $0 - \epsilon$ and two children of value $0.1$. Similarly $c_2$ is heapable below $a_5$ if and only if each sequence $\Delta(i, 4, 1)$ pairs off with and is heaped below the corresponding sequence $\Gamma(i - \epsilon, 1, 2)$. □

The centerpiece of our reduction, the *subset sequence* $b$, is comprised of $m$ subsequences representing the $m$ subsets in $\mathcal{C}$. For each subset $C_i = \{u_i, v_i, w_i\}$, let $u_i < v_i < w_i$ w.l.o.g. and let $b_i$ be the sequence of 18 values

$$b_i = (-1, i, 0), (-1, i, 1), (K, i, 1), (K, i, 0), (u_i, 0, 0), (v_i, 0, 0), (w_i, 0, 0),$$
$$\Delta(0, 1, 2), (L, i, 8), (L, i, 7), \ldots, (L, i, 1)$$

Now take $b = b_m, b_{m-1}, \ldots, b_1$. Claim 6 implies that if $x$ is completely heapable then $b$ must totally fit into the remaining *free* slots of the heap (i.e., the black area in Figure 1).

**Claim 8.** *If $x$ is completely heapable, then the $m$ roots of the complete binary trees comprising the tree slots must be the initial $(-1, i, 0)$ values from each of the $b_i$ subsequences.*

*Proof.* Observe that the $(-1, i, 0)$ values form a decreasing subsequence, and are too small for any of the singleton slots. They must therefore occupy space in the $m$ complete binary trees. By Claim 3, they mutually have no ancestor/descendant relationship, and must be in separate trees. But as they are the $m$ smallest values in $b$ they must occupy the $m$ *roots* of these trees. □

Claim 8 implies that the values of each $b_i$ must be slotted into a single binary tree in the black area of Figure 1 as well as some singleton slots. The following claim shows that the values occupying the singleton slots correspond to choosing the entire subset $C_i$ or not choosing it at all.

**Claim 9.** *If $x$ is completely heapable, then each $b_i$ sequence fills exactly 15 tree slots from a single complete binary tree and exactly 3 singleton slots. Furthermore, the 3 singleton values are either the three values $(u_i, 0, 0), (v_i, 0, 0), (w_i, 0, 0)$ or the three values $\Delta(0, 1, 2)$.*

*Proof.* By Claims 3 and 2, the $8m$ decreasing $L$ values must occupy level 4 (i.e. the final row of the black area in Figure 1). For a given subsequence $b_i$, Claim 8 implies that the suffix $(L, i, 8), \ldots, (L, i, 1)$ occupy the leaves of the binary tree rooted at $(-1, i, 0)$. As a consequence, we need to select a completely-heapable subsequence of length exactly 7 from the residual prefix of $b_i$ (prior to $(L, i, 8)$).

First, note that the first four values of $b_i$ must be included, as they cannot be placed elsewhere in the heap. Moreover, the orientation of these four values is forced: since $(K, i, 1)$ and $(K, i, 0)$ can only be parents of nodes of the form $(L, i, *)$, they must be placed at level two, with $(-1, i, 1)$ as their parent at level one.

Now consider $(u_i, 0, 0)$. If this value is included in the complete heapable subsequence, its location is forced to be the available child of the root $(-1, i, 0)$, and therefore both $(v_i, 0, 0)$ and $(w_i, 0, 0)$ must also be selected as its children (the zeroes in $\Delta(0, 1, 2)$ are too large to be eligible) to conclude the complete heapable subsequence. The three values of $\Delta(0, 1, 2)$ are necessarily exiled to slack slots in this case. Alternatively, if $(u_i, 0, 0)$ is not selected in the complete heapable subsequence, then the three nodes concluding the heapable subsequence must be $\Delta(0, 1, 2)$, since neither $(v_i, 0, 0)$ nor $(w_i, 0, 0)$ has two eligible children in the considered prefix of $b_i$. Therefore, the three values $(u_i, 0, 0)$, $(v_i, 0, 0)$, $(w_i, 0, 0)$ are exiled to slack slots in this case. □

The hardness result follows directly from the following lemma.

**Lemma 1.** $(Y, \mathcal{C})$ *contains an exact cover iff $x$ is completely heapable.*

*Proof.* For the if-direction, examine the complete heap produced by $x$. For each $b_i$ tree, use subset $C_i$ as part of the exact cover if and only if that tree includes $\Delta(0, 1, 2)$ in its entirety. By Claim 9 the set values from $C_i$ were all assigned to the set cover slots which we know enforces a set cover by Claim 7, so the union of our $n/3$ subsets is an exact cover.

For the only-if direction, for each subset $C_i$ in the exact cover, heap the subset sequence $b_i$ so that $(u_i, 0, 0)$, $(v_i, 0, 0)$, $(w_i, 0, 0)$ occupy set cover slots and the remaining 15 values occupy tree slots. Taken together, these fill up the $n$ set cover slots and $n/3$ of the complete binary trees. Heap the $m - n/3$ subset sequences not in the cover so as to exile triples of the form $\Delta(0, 1, 2)$, filling up the $3m - n$ slack slots and the remaining $m - n/3$ complete binary trees. Since the epilogue $c$ perfectly seals the frontier created by $b$, $x$ is completely heapable. □

□

## 3.3. A linear-time algorithm for complete heapability of 0-1 sequences.

When we restrict the problem of complete heapability to 0-1 values, the problem becomes tractable. The basic idea is that any completely heapable sequence of 0-1 values can be heaped into a canonical shape dependent only upon the number of 1s appearing in the sequence. After counting the number of 1s, we attempt to heap the sequence into the shape. If it fails, the sequence is not heapable.

Without loss of generality, let $x$ be a sequence of $n = 2^k - 1$ 0-1 values since we can always pad the end of $x$ with 1s without affecting its complete heapability. With 0-1 sequences, once a 1 is placed in the tree, only 1s may appear below it. Thus, in any valid solution tree $T$ for $x$, the nodes labelled with 1 form a forest $\mathcal{F}(T)$ of perfect binary trees. Let $V(T)$ be the set of nodes of $T$ that are labeled with 0 and fall on a path from the root of $T$ to the root of a tree in $\mathcal{F}(T)$. Note that the nodes in $V(T)$ form a binary tree. Let $y_1, \ldots, y_r$ be the nodes of $V(T)$ in the order they appear in $x$. If $y_i$ is a non-full node in $V(T)$ then let $\alpha(y_i)$ be the number of nodes appearing in the perfect trees of 1s of which $y_i$ is the parent. If $y_i$ is a full node then let $\alpha(y_i) = 0$. Now let $\beta(y_i) = \alpha(y_i) + \beta(y_{i-1})$ where $\beta(y_1) = \alpha(y_1)$. The values $\beta(y_1), \ldots, \beta(y_r)$ represent the cumulative number of 1s that the first $i$ nodes in $V(T)$ can absorb from $\mathcal{F}(T)$. That is, after inserting $y_1, \ldots, y_i$, we can add at most $\beta(y_i)$ of the 1s appearing in $\mathcal{F}(T)$.

Suppose $x$ has $m$ 1s in total and let $T^*$ be a perfect binary tree of height $k$ where the first $m$ nodes visited in a post-order traversal of $T^*$ are labelled 1 and the remainder of nodes are labelled 0. Note that the nodes labelled with 1 in $T^*$ form a forest $\mathcal{F}(T^*) = T_1^*, T_2^*, \ldots, T_z^*$ of $z$ perfect binary trees in descending order by height. Let $v_1, \ldots, v_m$ be the nodes of $\mathcal{F}(T^*)$ given by sequential pre-order traversals of $T_1^*, T_2^*, \ldots, T_z^*$. Let $u_1, \ldots, u_s$ be the nodes given by a pre-order traversal of $V(T^*)$. We build $T^*$ so that the first $s$ 0s appearing in $x$ are assigned sequentially to $u_1, \ldots, u_s$ and the $m$ 1s appearing in $x$ are assigned sequentially to $v_1, \ldots, v_m$.

**Lemma 2.** *$x$ is completely heapable if and only if $T^*$ is a valid solution for $x$.*

*Proof.* It's clear that if $T^*$ is a valid solution for $x$ then, by definition $x$ is completely heapable. Now, suppose $x$ is completely heapable. Then there exists a valid solution tree $T$. We show that whenever a 1 is added to $T$, we can also add a 1 to $T^*$. It should be clear that whenever a 0 is added to $T$ we can add a 0 to $T^*$.

Let $y_1, \ldots, y_r$ be the nodes of $V(T)$ in the order they appear in $x$. Note that $s \leq r$. This follows because $\mathcal{F}(T^*)$ has the fewest number of binary trees in any valid solution for $x$. One way to see this is by imagining each perfect tree of 1s as corresponding to one of the $2^i - 1$ terms in the (unique) polynomial decomposition of $m$ into $m = a_l(2^l - 1) + a_{l-1}(2^{l-1} - 1) + \cdots + a_1(2^1 - 1)$ where each coefficient $a_i$ is either 0 or 1 except for the final non-zero coefficient which may be 2. This is essentially an "off-by-one" binary representation of $m$. Thus, the perfect trees in $\mathcal{F}(T^*)$ have strictly decreasing heights except for, potentially, the shortest two trees which may have identical heights. It's clear that assigning the 0s in this order makes the largest number of 1 slots available as quickly as possible in any valid solution tree. Thus, for for $1 \leq j \leq s$ we have $\beta(u_j) \leq \beta(y_j)$. Therefore, anytime a 1 is placed in $T$, we can place a 1 in $T^*$.                                                                            $\square$

We're now prepared to prove the main theorem of this section.

---

**Algorithm 1** COMPLETE-HEAP $(x)$ where $x$ is a sequence of $n = 2^k - 1$ 0-1 values

---

1: $T^* \leftarrow$ perfect binary tree with $n$ nodes $u_1, \ldots, u_n$
2: $m \leftarrow$ number of 1s in $x$
3: $Q \leftarrow$ empty queue
4: $\mathcal{F}(T^*) = \{T_1^*, \ldots, T_z^*\} \leftarrow$ a forest of $z$ trees given by the first $m$ nodes  in a post-order traversal of $T^*$ and ordered by height
5: **for** $i \leftarrow 1$ **to** $z$ **do**
6:     $Q_i \leftarrow$ a queue of nodes given by a pre-order traversal of $T_i^*$
7: **end for**
8: $Q_0 \leftarrow$ a queue of $n - m$ nodes given by a pre-order traversal of $T^* - \mathcal{F}(T^*)$
9: **for** $i \leftarrow 1$ **to** $n$ **do**
10:   **if** $x_i = 0$ **then**
11:       $u \leftarrow$ DEQUEUE$(Q_0)$
12:       **if** $u$ is the parent of some tree $T_j^*$ in $\mathcal{F}(T^*)$ **then**
13:           dequeue the elements from $Q_i$ and enqueue them into $Q$
14:       **end if**
15:   **else**
16:       $u \leftarrow$ DEQUEUE$(Q)$
17:   **end if**
18:   **if** $u =$ NIL **then**
19:       **return**  "NOT HEAPABLE"
20:   **else**
21:       assign $x_i$ to $u$
22:   **end if**
23: **end for**
24: **return** $T^*$

---

**Theorem 3.** *Complete heapability of sequences of 0-1 values is decidable in linear time.*

*Proof.* Algorithm 1 provides a definition of Complete-Heap which we use to decide in linear time if $x$ is completely heapable. Initially, we build an unlabeled perfect binary tree of height $k$. We also count the number of 1s appearing in $x$. Both these operations take linear time. Next we identify where and in what order the 1s should be assigned and build a queue of nodes $Q_i$ for each tree $T_i^* \in \mathcal{F}(T^*)$. These operations take linear time in total since we can build the $T_i^*$ in one post-order traversal of $T^*$ and each $Q_i$ can be built from a single pre-order traversal of $T_i^*$. We also identify where and in what order the 0s should be assigned to $T^*$ and enqueue these nodes in $Q_0$.

Now we simply try and assign each value in $x$ to the appropriate node in $T^*$ if it is available. The idea is that once the parent of tree $T_i^*$ gets labeled with a 0, then the nodes in $Q_i$ are available for assignment. We can mark these parent nodes ahead of time to ensure our algorithm runs in linear time. If $Q$ ever runs dry of nodes, then we don't have enough 0s to build the frontier necessary to handle all the 1s, so $x$ is not completely heapable. On the other hand, if we terminate without exhausting $Q$, then the sequence is completely heapable. The correctness of the algorithm follows immediately from Lemma 2. $\square$

## 4. Heapable Subsequences

In this section, we focus on the case where the sequence corresponds to a random permutation. There are three standard models in this setting. In the first, the sequence is known to be a permutation of the numbers from 1 to $n$, and each element is a corresponding integer. Let us call this the permutation model. In the second case, the sequence is again known to be a permutation of $[1, n]$, but when an element arrives one is given only its ranking relative to previous items. Let us call this the relative ranking model. In the third, the sequence consists of independent uniform random variables on $(0, 1)$. Let us call this the uniform model. All three models are equivalent in the offline setting, but they differ in the online setting, where the relative ranking model is the most difficult.

We first show that the longest heapable subsequence in any of these models, has length $(1-o(1))n$ with high probability, and in fact such subsequences can even be found online. For simplicity we first consider the offline case for the uniform model. We then show how to extend it to the online setting and to the relative ranking model. (As the permutation model is easier, the result follows readily for that model as well.) We note that we have not attempted to optimize the $o(1)$ term. Finding more detailed information regarding the distribution of the LHS in these various settings is an open problem.

**Theorem 4.** *In the uniform model, the longest heapable subsequence has length $(1 - o(1))n$ with high probability.*

*Proof.* We break the proof into two stages. We first show that we can obtain an LHS of length $\Omega(n)$ with high probability. We then bootstrap this result to obtain the theorem.

Let $A_1$ be the subsequence consisting of the elements with scores less than $1/2$ in the first $n/2$ elements. With high probability the longest increasing subsequence of $A_1$ is of length $\Omega(\sqrt{n})$. Organize the elements from the LIS of $A_1$ into a heap, with $F = \Omega(\sqrt{n})$ leaf nodes.

Now let $A_2$ be the subsequence consisting of the elements with scores greater than $1/2$ in the last $n/2$ elements. Starting with the heap obtained from $A_1$, we perform the greedy algorithm for the elements of $A_2$ until the first time we cannot place an element. Our claim is that with high probability a linear number of elements are placed before this occurs. Consider the $F$ subheaps, ordered by their root element in decreasing order. In order not to be able to place an element, we claim that we have seen a decreasing subsequence of $F$ elements in $A_2$. This follows from the same argument regarding the length of the LIS derived from patience sorting. Specifically, each time an element was placed on a subheap other than the first, there must be a corresponding larger element placed previously on the previous subheap. Hence, when we cannot place an element, we have placed at least one element on each subheap, leading to a chain corresponding to a decreasing subsequence of $F$ elements. As $F = \Omega(\sqrt{n})$, with high probability such a subsequence does not appear until after successfully placing $\Omega(n)$ elements of $A_2$.

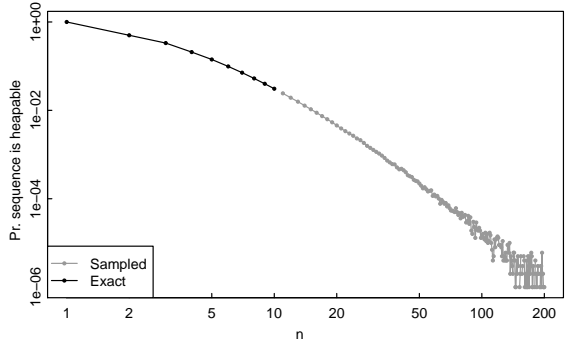

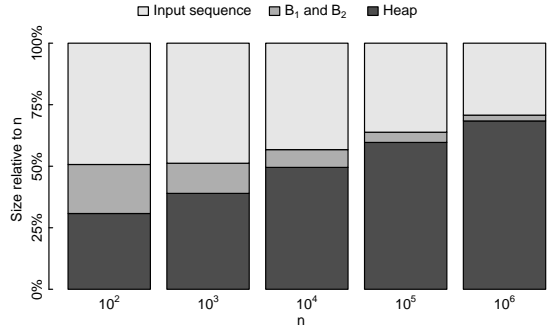

FIGURE 3. The probability that a random permutation of $n$ numbers is heapable as $n$ varies. For values of $n$ up to 10 the probabilities are exact; for larger values of $n$ they are estimated from a set of $10! \approx 4 * 10^6$ sample permutations.

FIGURE 4. The size of the heap found using the algorithm described in Theorem 4, as well as the joint length of subsequences $B_1$ and $B_2$, both with respect to the length of the input sequence $n$.

Given this result, we now prove the main result. Let $B_1$ be the subsequence consisting of the elements less than $n^{-1/8}$ in the first $n^{7/8}$ elements. With high probability there are $\Omega(n^{3/4})$ elements in $B_1$ using standard Chernoff bounds, and hence by the previous paragraphs we can find an LHS of $B_1$ of size $\Omega(n^{3/4})$. Now let $B_2$ be the subsequence consisting of the elements greater than $n^{-1/8}$ in the remaining $n - n^{7/8}$ elements. We proceed as before, performing the greedy algorithm for the elements of $B_2$ until the first time we cannot place an element. For the process to terminate before all elements of $B_2$ having been placed, $B_2$ would have to have an LDS of length $\Omega(n^{3/4})$, which does not occur with high probability.                                                                      □

We implemented the algorithm described in Theorem 4 and applied it to a range of sequences of increasing size. Figure 4 displays the size of the resulting heap (averaged over 1000 iterations for each value of $n$) relative to the length of the original sequence, $n$.

The proof extends to the online case.

## 4.1. The case of random permutations.

**Corollary 1.** *In the uniform model, a heapable subsequence of length $(1 - o(1))n$ can be found online with high probability.*

*Proof.* We use the fact that there are online algorithms that can obtain increasing subsequences of length $\Omega(\sqrt{n})$ in random permutations of length $n$ [7]. Using such an algorithm on $A_1$ as above gives us an appropriate starting point for using the greedy algorithm, which already works in an online fashion, on $A_2$, to find an increasing subsequence of length $\Omega(n)$ with high probability. We can then similarly extend the proof as in Theorem 4 to a sequence of length $(1 - o(1))n$ using the subsequences $B_1$ and $B_2$ similarly.                                                                      □

There are various ways extend these results to the relative ranking model. For the offline problem, we can treat the first $\epsilon n$ elements as a guide for any constant $\epsilon > 0$; after seeing the first $\epsilon n$ elements, perform the algorithm for the uniform model for the remaining $(1-\epsilon)n$ elements, treating an element as having a score less than $1/2$ if it is ranked higher than half of the initial $\epsilon n$ elements and greater than $1/2$ otherwise. The small deviations of the median of the sample from the true median will not affect the asymptotics of the end result. Then, as in Theorem 4, bootstrap to obtain an algorithm that finds a sequence of length $(1 - o(1))n$.

For the online problem, we are not aware of results giving bounds on the length of the longest increasing (or decreasing) subsequence when only relative rankings are given, although it is not difficult to obtain an $\Omega(\sqrt{n})$ high probability bound given previous results. For example, one could similarly use the above approach, using the first $\epsilon n$ elements as a guide to assign approximate $(0, 1)$ values to remaining elements, and then use a variation of the argument of Davis (presented in [7][Section 7]) to obtain a longest increasing subsequence on the first half of the remaining elements of size $\Omega(\sqrt{n})$.

We describe a more direct variation. Order the first $\epsilon n$ elements, and split the lower half of them by rank into $\sqrt{n}$ subintervals. Now consider next $(1-\epsilon)n/2$ elements. Split them, sequentially, into $\sqrt{n}$ subgroups; if the $i$th subgroup of elements contains an element that falls in the $i$th subinterval, put it in our longest increasing subsequence. Note that this can be done online, and for each subinterval the probability of obtaining an element is a constant. Hence the expected size of the longest increasing subsequence obtained this way is $\Omega(\sqrt{n})$, and a standard martingale argument can be used to show that in fact this holds with high probability. Then, as before we can show that in the next $(1-\epsilon)n/2$ elements, we add $\Omega(n)$ elements to our heap with high probability using the greedy algorithm. As before, this gives the first part of our argument, which can again be bootstrapped.

**Corollary 2.** *In the relative ranking model, a heapable subsequence of length $(1 - o(1))n$ can be found both offline and online with high probability.*

We now turn our attention to the problem of finding the longest completely heapable subsequence in the uniform and relative ranking models, as well as the associated online problems. For convenience we start with finding completely heapable subsequences online in the uniform model, and show that we can obtain sequence of length $\Omega(n)$ with high probability. Our approach here is a general technique we call *banding*; for the $i$th level of the tree, we only accept values within a band $(a_i, b_i)$. We chose values so that $a_1 < b_1 = a_2 < b_2 = a_3 \ldots$, that the bands are disjoint and naturally yield the heap property. Obviously this gives that the LCHS is $\Omega(n)$ with high probability as well. We note no effort has been made to optimize the leading constant in the $\Omega(n)$ term in the proof below.

**Theorem 5.** *In the uniform model, a completely heapable subsequence of length $\Omega(n)$ can be found online with high probability.*

*Proof.* As previously, we can find an LIS of size $\Omega(\sqrt{n})$ online within the first $n/2$ elements restricted to those with value less than $1/2$. This will give the first $(\log n)/2 - c_1$ levels of our heap, for some constant $c_1$.

We now use the banding approach, filling subsequent levels sequentially. Suppose from the LIS that our bottom level has $t_0$ nodes. Consider the next $u_1$ elements, and for the next level use a band of size $v_1$, which in this case corresponds to the range $(1/2, 1/2 + v_1)$. We need $t_1 = 2t_0$ elements to fill the next level. Note that if we choose for example $u_1 v_1 = 2t_1 = 4t_0$, we will be safe, in that Chernoff bounds guarantee we obtain enough elements to fill the next level. We let $u_1 = 2\sqrt{t_0}n^{1/2}$ and $v_1 = u_1/n$.

For each subsequent level we will need twice as many items, so generalizing for the $i$th level after the base we have $t_i = 2^i t_0$, and we can consider the next $u_i = (\sqrt{2})^{i+1}\sqrt{t_0}n^{1/2}$ elements using a band range of size $v_i = u_i/n$. We continue this for $L$ levels. As long as $\sum_{i=1}^{L} u_i \leq n/2$ and $\sum_{i=1}^{L} v_i \leq 1/2$, the banding process can fill up to the $L$th level with high probability. As the sums are geometric series, it is easy to check that we can take $L = (\log n)/2 - c_2$ for some constant $c_2$ (which will depend on $t_0$). This gives the result, and the resulting tree now has $\log n - c_1 - c_2$ levels, corresponding to $\Omega(n)$ nodes. □

We implemented the algorithm described in Theorem 5 and applied it to a range of sequences of increasing size. For each sequence size, Figure 5 displays the average number of levels in the resulting perfect heap. We verify that the number of elements of the resulting heap grows linearly to the length of the original sequence, as expected.

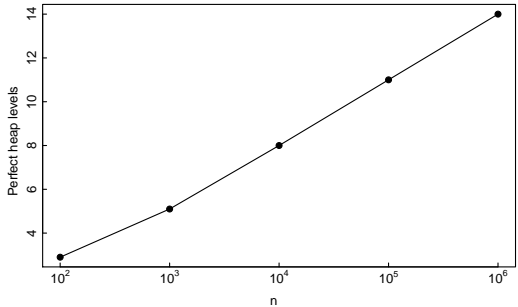

FIGURE 5. The number of levels in a perfect heap constructed using the algorithm described in Theorem 5 as $n$ varies. Note the logarithmic scale of the $x-$axis.

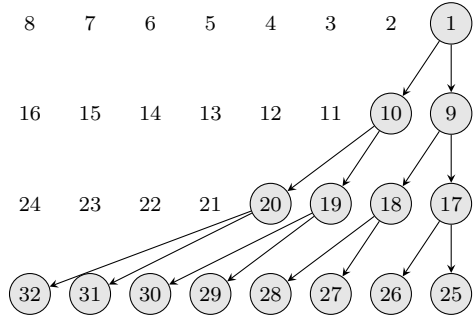

FIGURE 6. An illustration of Theorem 6 for $n = 32$. The elements are ordered left-to-right, top-to-bottom. For example 8 precedes 7 and 1 precedes 16. A representative longest increasing heapable subsequence is highlighted.

We can similarly extend this proof to the relative ranking case. As before, using the first $\epsilon n$ elements as guides by splitting the lower half of these elements into $\sqrt{n}$ regions, we can obtain an increasing sequence of size $\Omega(\sqrt{n})$ to provide the first $(\log_n)/2 - c_1$ levels of the heap. We then use the banding approach, but instead base the bands on upper half of first $\epsilon n$ elements in the natural way. That is, we follow the same banding approach as in the uniform model, except when the band range is $(\alpha, \beta)$ in the uniform model, we take elements with rankings that fall between the $\lceil \alpha \epsilon n \rceil$th and $\lfloor \beta \epsilon n \rfloor$th of the first $\epsilon n$ elements. It is straightforward to show that with high probability this suffices to successfully fill an additional $(\log n)/2 - c_2$ levels, again given a completely heapable subsequence of length $\Omega(n)$.

Again, for all of these variations, the question of finding exact assymptotics or distributions of the various quantities provides interesting open problems.

4.2. **Longest increasing and decreasing heapable subsequences.** Because the longest heapable subsequence problem is a natural variation of the longest increasing subsequence problem, and the latter has given rise to many interesting combinatorial problems and mathematical connections, we expect that the introduction of these ideas will lead to many interesting problems worth studying. For example, as we have mentioned, one of the early results in the study of increasing subsequences, due to Erdös and Szekeres, is that every sequence of $n^2 + 1$ distinct numbers has either an increasing or decreasing subsequence of length $n+1$ [4]. One could similarly ask about the longest increasing or decreasing heapable subsequence within a sequence. We have the following simple upper bound; we do not know whether it is tight.

**Theorem 6.** *There are sequences of $n$ elements such that the longest increasing or decreasing heapable subsequence is upper bounded by $O(n/\log n)$.*

*Proof.* In fact we can show something stronger; there are sequences such that the longest increasing heapable subsequence and the longest decreasing subsequence have length $O(n/\log n)$. Consider the following construction: we begin by splitting the sequence of $n$ elements into $B$ equally sized blocks. Each block is a decreasing subsequence, and the subsequences are in increasing order, as illustrated in Figure 6. It can be easily seen that the longest decreasing subsequence has length $n/B$. For the longest increasing heapable subsequence, note that our optimal choice is to take one element from the first block, two from the next block, and so on so forth. We want to select an appropriate value for $B$ so that the last block is the last full level of our increasing heap. The number of heap elements is then $2^B - 1$. Setting $2^B - 1$ and $n/B$ equal we have $B(2^B - 1) = n$, which for large $n$ is approximated by $B2^B = n$. Recall that the solution to this equation is $B = W(n)$ where $W$ is the Lambert $W$ function. The latter has no closed form but a reasonable approximation is $\log n - \log \log n$, so asymptotically we can arrange a bound of $O(n/\log n)$.  □

## 5. Open Problems

Besides finding tight bounds for the problem in the previous section, there are several other interesting open questions we have left for further research.

- Is there an efficient algorithm for finding the longest heapable subsequence, or is it also NP-hard? If it is hard, are there good approximations?
- For binary alphabets, we have shown complete heapability can be decided in linear time, while for permutations on $n$ elements, the problem is NP-hard. What is the complexity for intermediate alphabet sizes?
- What is the probability that a random permutation is heapable – either exactly, or asymptotically?
- Can we find the exact expected length or the size distribution of the longest heapable subsequence of a random permutation? The longest completely-heapable subsequence?
- The survey of Aldous and Diaconis [1] for LIS shows several interesting connections between that problem and patience sorting, Young tableaux, and Hammersley's interacting particle system. Can we make similar connections to these or other problems to gain insight into the LHS of sequences?

We expect several other combinatorial variations to arise.

There are also many open problems relating to our original motivation: viewing this process as a variation of the hiring problem. For example, we can consider the quality of a hiring process as corresponding to some function of the ranking or scores of the people hired, as in [3]. Here we have focused primarily on questions of maximizing the length of the sequence, or equivalently the number of people hired. More general reward functions, such as penalizing unfilled positions or allowing for errors such as an employee being more qualified than their boss in the hierarchy tree, seem worthy of further exploration.